\documentclass{IEEEtran}
\usepackage[T1]{fontenc}
\usepackage{cite}
\usepackage{amsmath,amssymb,amsfonts}
\usepackage{algorithmic}
\usepackage{graphicx}
\usepackage{textcomp}
\usepackage{xcolor}
\usepackage{setspace}
\usepackage{enumitem}

\def\BibTeX{{\rm B\kern-.05em{\sc i\kern-.025em b}\kern-.08em
		T\kern-.1667em\lower.7ex\hbox{E}\kern-.125emX}}
\usepackage{subfigure}

\newtheorem{my_lemma}{Lemma}

\title{Diversity Analysis of Multi-Aperture UWOC System over EGG Channel with Pointing Errors}

\author{Ziyaur Rahman, ~\IEEEmembership{Student Member,~IEEE}, Ankur Bansal, ~\IEEEmembership{Member,~IEEE} and S.~M.~ Zafaruddin, ~\IEEEmembership{Senior Member,~IEEE}
	\thanks{
		
		This work was supported in part by the 
		Science and Engineering Research Board (SERB), India under MATRICS Grant MTR/2021/000890 and Start-up Research Grant SRG/2019/002345.
			
		Ziyaur Rahman and  S.~M.~ Zafaruddin are  with  the Department of Electrical and Electronics Engineering, Birla Institute of Technology and Science Pilani at Pilani,  Rajasthan, India 	(email: p20170416@pilani.bits-pilani.ac.in, syed.zafaruddin@pilani.bits-pilani.ac.in) 
		
		Ankur Bansal is with the Department of Electrical Engineering, IIT Jammu, Jagti Campus, Jammu \& Kashmir-181221, India
		(email: ankur.bansal@iitjammu.ac.in).}
	
}

\begin{document}
	\maketitle
	\begin{abstract}
			The existing single-aperture reception for underwater wireless optical communication (UWOC) is insufficient to deal with oceanic turbulence caused by the combined effect of temperature gradient and air bubbles. This paper analyzes the performance of multi-aperture reception for UWOC under channel irradiance fluctuations characterized by the mixture exponential–generalized gamma (EGG) distribution. We analyze the system performance by employing both selection combining (SC) and maximum ratio combining (MRC) receivers. In particular, we derive the exact outage probability expression for the SC-based multi-aperture UWOC receiver and obtain an upper bound on the outage probability for the MRC-based multi-aperture UWOC receiver. With the help of the derived results, we analytically obtain the diversity order of the considered multi-aperture UWOC systems  outperforming the single-aperture-based system by a factor of the number of multi-aperture elements. The simulation  results validate the accuracy of derived analytical expressions and demonstrate the performance improvement of the  multi-aperture UWOC system compared to a single-aperture-based system.
	\end{abstract}
	
	\begin{IEEEkeywords}
		EGG model, MRC, oceanic turbulence, outage probability,  performance analysis, pointing errors, SC.
	\end{IEEEkeywords}

	\section{Introduction}
Underwater wireless optical communication (UWOC)  is a potential technology for  submarine  and oceanic scientific applications \cite{Gussen2016, Kaushal2016, Zeng2017}. The UWOC system can provide higher throughput due to enormous contiguous bandwidth available in the optical band compared with the acoustic and radio frequency (RF). However, the underwater link suffers from  signal attenuation due to oceanic path loss, turbulence, and pointing errors.  Oceanic turbulence is the effect of random variations in the refractive index of the  UWOC channel caused by  random variations in the water temperature, salinity, and air bubbles. It is  desirable to improve the UWOC system performance over various underwater channel impairments for a better quality of service.

A better performance assessment of UWOC requires accurate characterization of turbulence-induced fading under various underwater conditions. In \cite{Jamali2018} \cite{Zedini2019}, the authors proposed a holistic experimental view on the statistical characterization of oceanic turbulence in UWOC systems, considering the effect of the temperature gradient, salinity, and air bubbles.  The authors in \cite{Zedini2019} used experimental data to propose the mixture exponential-generalized Gamma (EGG) distribution for oceanic turbulence caused by air bubbles and   temperature gradient for UWOC channels perfectly matches the measured   data collected under different channel conditions ranging from weak to strong turbulence. 
    
The  unifying nature of the EGG model  depicting various oceanic channel conditions has sparked research interest in analyzing the UWOC system performance  
\cite{Zedini2020, Lei2020, Li2020, Yang2021, Li2021, Anees2019, Phookan2020, Rahman2022_direct, Le2022, Lou2022, Rahman2022_unified}.  In \cite{Zedini2020}, the ergodic capacity, outage probability, and the average BER  over EGG turbulence channels for dual-hop UWOC systems have been analyzed. The  mixed dual-hop mixed RF-UWOC system over EGG model has been extensively investigated in \cite{Lei2020, Li2020, Yang2021, Li2021}. The authors in \cite{Anees2019}  derived an  expression for average BER considering where the RF link is modeled by Nakagami-m fading. The average symbol-error-rate (SER) for the UWOC system over EGG distributed turbulence channel using $2 \times 1$ Alamouti-based space-time block code has been provided in \cite{Phookan2020}. A direct air-to-underwater optical wireless communication system performance is provided in \cite{Rahman2022_direct}. The authors in \cite{Le2022} derived tight mathematical expressions of outage probability and BER for multi-hop UWOC systems. The performance of the multi-layer UWOC system is discussed in \cite{Lou2022, Rahman2022_unified, Das2022}.

In the existing literature, the authors have considered the performance analysis for the single-aperture UWOC link over different oceanic turbulence conditions. To the best of the authors’ knowledge, the multi-aperture UWOC system with various combining receivers has not been analyzed, which motivates us to find the performance of the multi-aperture UWOC system under the combined effect of oceanic turbulence and pointing errors. In this paper, we analyze the performance of a multi-aperture UWOC system under the combined impact of oceanic turbulence and pointing errors using selection combining (SC) and maximum ratio combining (MRC) diversity schemes. It should be emphasized that deriving statistical results for the multi-aperture system is challenging due to the sum of two Meijer's-G functions modeling the density function for the $i$-th aperture.  The major contributions of the proposed work are summarized as follows:
\begin{itemize}
	\item  We derive the analytical expressions of probability density function (PDF) and cumulative distribution function (CDF) of the instantaneous signal-to-noise
	ratios (SNR) at the multi-aperture-based receiver under both the MRC and SC schemes.
	
	\item Under the SC scheme, we obtain the exact outage probability expression for the considered UWOC system over EGG oceanic turbulence and pointing errors. Moreover, we derive an upper bound on outage probability for MRC based receiver under the effect of pointing errors.
	
	\item Further, we analyze the considered UWOC system under high SNR conditions and obtain the asymptotic expressions of outage probability for both MRC and SC receivers. The diversity order of the considered system is obtained analytically under MRC and SC techniques.
	
	\item The accuracy and the tightness of the derived analytical results are verified through simulations. The results demonstrate the performance improvement of the considered multi-aperture UWOC system compared to a single-aperture-based UWOC system.
\end{itemize}
	
	\begin{figure}[tp]	
		\centering
		\includegraphics[scale=0.7]{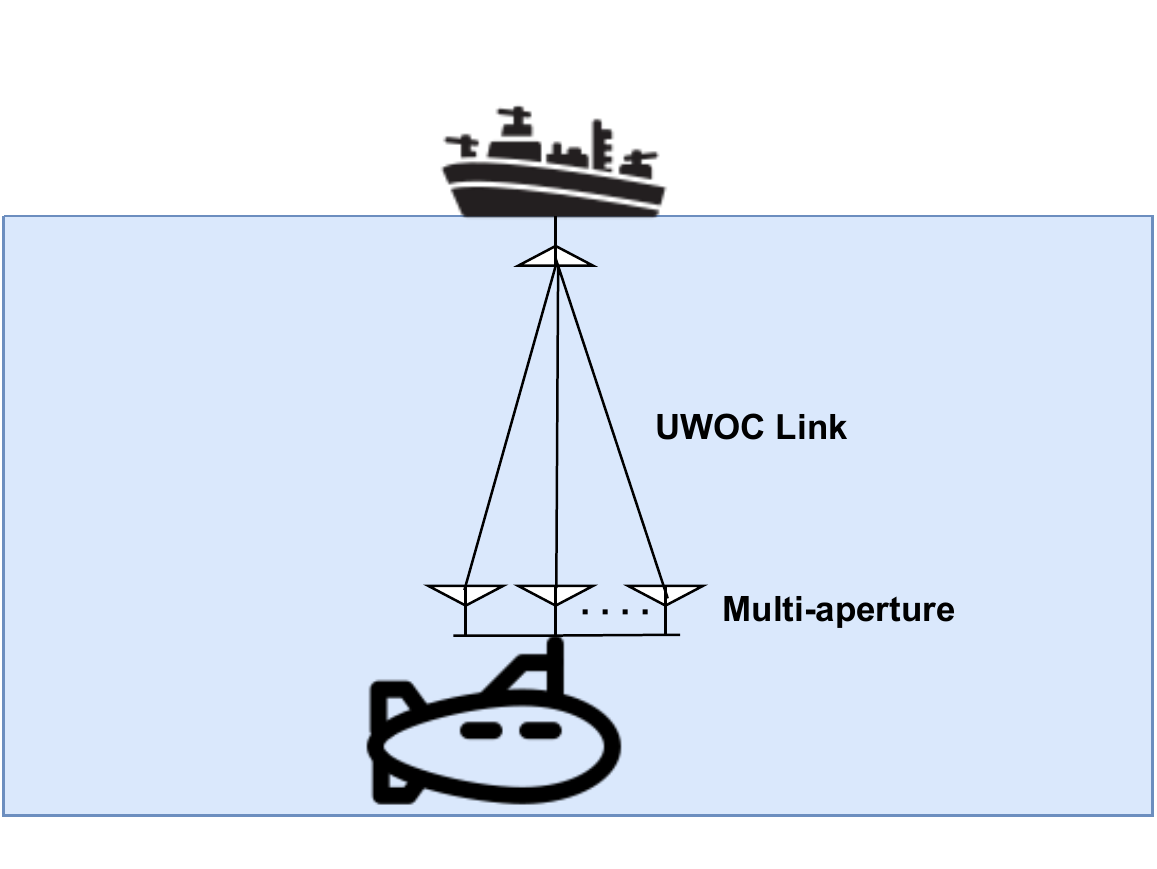}
		\caption{ Schematic diagram  for multi-aperture UWOC system.}
		\label{system_model}	
	\end{figure}
	
	\section{System Model}
	We consider a multi-aperture diversity scheme (MRC and SC) at destination for UWOC systems over  EGG fading channel  with the transceiver antenna misalignment (as shown in Fig.~\ref{system_model}). The received signal $y_i$ at the $i$ receiver aperture using non-coherent intensity modulation/direct detection (IM/DD) scheme is given by: 
	\begin{eqnarray}
	&y_i =h_{l,i}h_{t,i} h_{p,i} s + w_i
	\label{eq:received signal}
	\end{eqnarray}
	where $h_{l,i}=e^{-\alpha l}$ is the oceanic path loss coefficient (with link distance $l$ (in \rm{m}) and extinction attenuation coefficient $\alpha$), $h_{t,i}$ is the fading channel coefficient between the source and $i$-th detector aperture, $h_{p,i}$ is the misalignment errors (i.e., pointing errors) between transceiver, $s$ is the transmitted signal, and $w_i$ is the additive white Gaussian noise (AWGN) with variance $\sigma_w^2$.	Using the on-off-keying (OOK) signaling, the instantaneous SNR  is given by 
	\begin{equation}
	\gamma_i = \frac{P^2_t h_{l,i}^2 h_{t,i}^2h_{p,i}^2}{\sigma^2_w} = \gamma_0 h_{t,i}^2h_{p,i}^2,
	\label{eq:inst.snr}
	\end{equation}
	where $P_t$ is the transmitted optical  power such that $s \in \{0,P_t \}$ and 
		$\gamma_0= \frac{P^2_t h_{l,i}^2}{\sigma^2_w}$.

 We consider recently \cite{Zedini2019} proposed EGG distribution (i.e., the combined exponential and generalized Gamma function) to model the oceanic turbulence. The probability density function (PDF) of the EGG distributed fading channel is given as \cite{Zedini2019}: 
\begin{flalign}
f_{h_{t,i}}(x) = \frac{\omega_i}{\lambda_i}\exp(-\frac{x}{\lambda_i}) + (1-\omega_i)\frac{c_i x^{a_ic_i-1}}{b_i^{a_ic_i}}\frac{\exp(-(\frac{x}{b_i})^{c_i})}{\Gamma(a_i)}
\label{pdf_uw}
\end{flalign}
where $\omega_i$ is the mixture coefficient of the distributions (i.e., $0<\omega_i<1$), $\lambda_i$, $a_i$, $b_i$, and $c_i$ are the exponential distribution parameters.

 The PDF and CDF of SNR over EGG fading channel with pointing errors are given as \cite{Yang2021}:
 \begin{eqnarray}
 &f_{\gamma_i}(\gamma) =  \frac{\omega_i \rho_i^2 }{2 \gamma} G_{1,2}^{2,0}\left(\begin{array}{c}\rho_i^2+1\\1,\rho_i^2\end{array}\left|\frac{1}{\lambda_i A_i} \left(\sqrt{\frac{\gamma}{\gamma_0}}\right)\right.\right) \nonumber\\&+ \frac{(1-\omega_i)\rho_i^2}{2 \Gamma(a_i) \gamma} G_{1,2}^{2,0}\left(\begin{array}{c}\frac{\rho_i^2}{c_i}+1\\a_i,\frac{\rho_i^2}{c_i}\end{array}\left|\frac{1}{b_i^{c_i} A_i^{c_i}}\left(\sqrt{\frac{\gamma}{\gamma_0}}\right)^{c_i}\right.\right)
 \label{pdf_uw_pe}
 \end{eqnarray}
 
 \begin{eqnarray}
 &F_\gamma(\gamma_i) = \omega_i\rho_i^2 G_{2,3}^{2,1}\left(\begin{array}{c}1,\rho_i^2+1\\1,\rho_i^2,0\end{array}\left|\frac{1}{\lambda_i A_i}\left(\sqrt{\frac{\gamma}{\gamma_0}}\right)\right.\right) \nonumber\\&+\frac{(1-\omega_i)\rho_i^2}{c_i \Gamma(a_i)}G_{2,3}^{2,1}\left(\begin{array}{c}1,\frac{\rho_i^2}{c_i}+1\\a_i,\frac{\rho^2}{c_i},0\end{array}\left|\frac{1}{b_i^{c_i} A_i^{c_i}}\left(\sqrt{\frac{\gamma}{\gamma_0}}\right)^{c_i}\right.\right)
 \label{cdf_uw_pe}
 \end{eqnarray}
where $A_i=\mbox{erf}(\upsilon)^2$ with $\upsilon=\sqrt{\pi/2}\ r/\omega_{z}$, $r$ is the aperture radius and $\omega_{z}$ is the beam width, and $\rho_i = {\frac{\omega_{{z}_{\rm eq}}}{2 \sigma_{s}}}$ with  $\omega_{{z}_{\rm eq}}$ as the equivalent beam width at the receiver and $\sigma^2_{s}$ as the variance of pointing errors displacement characterized by the horizontal sway and elevation \cite{Farid2007}.

\section{Performance analysis}
In this section, we develop a statistical analysis for the multi-aperture UWOC system over EGG-distributed oceanic turbulence with pointing errors. We also derive analytical expressions for the PDF and CDF to analyze the performance of multi-aperture UWOC system using MRC and SC diversity scheme.

\subsection{Maximum Ratio Combining}
Maximum ratio combining is one of the most popular diversity schemes, where the sum of statistical characterization is required \cite{Zheng2019}.
The PDF and CDF of SNR of $N$ aperture based UWOC channel using the MRC technique at the 
receive $\gamma_{\rm MRC}=\sum_{i=1}^{N}\gamma_i$ is quite complicated. Thus, we use an upper bound, which can be given as $\gamma_{\rm MRC}\leq N\gamma_N$, where  $\gamma_N=\prod_{i=1}^{N}\gamma^{\frac{1}{N}}_{i}$.  We use the inverse Mellin transform to find the PDF of $\gamma_N$. If $\mathbb{E}[X^r]$ denotes the $r$-th moment, where $\mathbb{E}[\cdot]$ denotes the expectation operator, then the inverse Mellin transform results the PDF of a random variable $X$ as \cite{bhardwaj2021performance}
\begin{equation}
f_{\gamma_N}(x)=\frac{1}{2\pi \jmath}\frac{1}{x}\int_{\cal{L}}^{}\prod_{i=1}^{N}\mathbb{E}[\gamma^{\frac{r}{N}}_i]x^{-r} dr
\label{eq:Nth_order_with_tnverse_Mellin}
\end{equation}
where the $r$-th moment is given as
\begin{eqnarray}
&\mathbb{E}[\gamma _i^{\frac{r}{N}}]=\int_{0}^{\infty}
x^{\frac{r}{N}} f_{\gamma_i}(x)dx
\label{eq:nth_order_moment_of_hi}
\end{eqnarray}

We find the PDF and CDF of SNR  for an $N$ aperture based UWOC system over EGG turbulence channel with pointing errors in the following Lemma.

\begin{my_lemma}
	The PDF and CDF of SNR of $N$-aperture based UWOC system over an EGG fading channel with pointing errors are given as:
	\begin{flalign}
	&f_{\gamma_N}(\gamma)=\prod_{i=1}^{N} \frac{\omega_i}{\gamma}H_{N,2N}^{2N,0}\left[\begin{matrix} V_1 \\ V_2 \end{matrix} \bigg| \prod_{i=1}^{N}\left(\frac{1}{\gamma_0\lambda^2_{i} A^2_{i}}\right)^{\frac{1}{N}}\gamma\right]\nonumber\\& +\prod_{i=1}^{N} \frac{(1-\omega_i)}{\gamma c_i \Gamma(a_i)}H_{N,2N}^{2N,0}\left[\begin{matrix} V_3 \\ V_4 \end{matrix} \bigg| \prod_{i=1}^{N}\left(\frac{1}{\gamma_0b^2_{i}A^2_{i}}\right)^{\frac{1}{N}}\gamma\right]
	\label{pdf_snr_mrc_uw_pe}
	\end{flalign}
	
	\begin{flalign}
	&F_{\gamma_N}(\gamma)=\prod_{i=1}^{N} \omega_iH_{N+1,2N+1}^{2N,1} \left[\begin{matrix} (1,1), V_1 \\ V_2, (0,1) \end{matrix} \bigg| \prod_{i=1}^{N}\left(\frac{1}{\gamma_0\lambda^2_{i} A^2_{i}}\right)^{\frac{1}{N}}\gamma\right]\nonumber\\& + \prod_{i=1}^{N} \frac{(1-\omega_i)}{ c_i \Gamma(a_i)}H_{N+1,2N+1}^{2N,1}\left[\begin{matrix} (1,1),  V_3 \\ V_4, (0,1) \end{matrix} \bigg| \prod_{i=1}^{N}\left(\frac{1}{\gamma_0b^2_{i}A^2_{i}}\right)^{\frac{1}{N}}\gamma\right]
	\label{cdf_snr_mrc_uw_pe}
	\end{flalign}
	where $V_1=\{(\rho_{1}^2+1,\frac{2}{N}), \cdots, (\rho_{N}^2+1,\frac{2}{N})\}$, $V_2=\{(1,\frac{2}{N}), \cdots, (1,\frac{2}{N})_N, (\rho_{1}^2,\frac{2}{N}), \cdots, (\rho_{N}^2,\frac{2}{N})\}$, $V_3=\{(\frac{\rho_{1}^2}{c_{1}}+1,\frac{2}{c_{1}N}), \cdots, (\frac{\rho_{N}^2}{c_{N}}+1,\frac{2}{c_{N}N})\}$ and $V_4=\{(a_{1},\frac{2}{c_{1}N}), \cdots, (a_{N},\frac{2}{c_{N}N}), (\frac{\rho_{1}^2}{c_{1}},\frac{2}{c_{1}N}), \cdots, (\frac{\rho_{N}^2}{c_{N}},\frac{2}{c_{N}N})\}$
\end{my_lemma}

\begin{IEEEproof}
	Substituting \eqref{pdf_uw_pe} in \eqref{eq:nth_order_moment_of_hi}, we get the $r$-th order moment of SNR as:
	
	\begin{eqnarray}
	& \mathbb{E}[\gamma_i^{\frac{r}{N}}] =\int_{0}^{\infty}
	\gamma^{\frac{r}{N}} \bigg[\frac{\omega_i \rho_i^2 }{2 \gamma} G_{1,2}^{2,0}\left(\begin{array}{c}\rho_i^2+1\\1,\rho_i^2\end{array}\left|\frac{1}{\lambda_i A_i} \left(\sqrt{\frac{\gamma}{\gamma_0}}\right)\right.\right) \nonumber \\&+ \frac{(1-\omega_i)\rho_i^2}{2 \Gamma(a_i) \gamma} G_{1,2}^{2,0}\left(\begin{array}{c}\frac{\rho_i^2}{c_i}+1\\a_i,\frac{\rho_i^2}{c_i}\end{array}\left|\frac{1}{b_i^{c_i} A_i^{c_i}}\left(\sqrt{\frac{\gamma}{\gamma_0}}\right)^{c_i}\right.\right)\bigg]d\gamma
	\label{r_th_momt_snr_uw}
	\end{eqnarray}
	
	To solve \eqref{r_th_momt_snr_uw}, we substitute $\sqrt{\frac{\gamma}{\gamma_0}}=t$ in the first term and then $t^{c_i}=u$ in the second term and applying the identity \cite[eq. 07.34.21.0009.01]{Wolfram}, we get $\mathbb{E}[\gamma_i^r]$ as:
	
	\begin{eqnarray}
	& \mathbb{E}[\gamma_i^{\frac{r}{N}}] = \omega_i \rho_i^2 \gamma_0^{\frac{r}{N}} \left(\frac{1}{\lambda_i A_i}\right)^{-\frac{2r}{N}}\frac{\Gamma\left(\frac{2r}{N}+1\right)\Gamma\left(\frac{2r}{N}+\rho_i^2\right)}{\Gamma\left(\frac{2r}{N}+(\rho_i^2+1)\right)}\nonumber \\&+ \frac{(1-\omega_i)\rho_i^2}{c_i\Gamma(a_i)}\gamma_0^{\frac{r}{N}} \left(\frac{1}{b_iA_i}\right)^{-\frac{2r}{N}} \frac{\Gamma\left(\frac{2r}{c_iN}+a_i\right)\Gamma\left(\frac{2r}{c_iN}+\frac{\rho_i^2}{c_i}\right)}{\Gamma\left(\frac{2r}{c_iN}+\left(\frac{\rho_i^2}{c_i}+1\right)\right)}
	\label{r_th_momt_snr_uw_1}
	\end{eqnarray}
	
Substituting \eqref{r_th_momt_snr_uw_1} in \eqref{eq:Nth_order_with_tnverse_Mellin} and applying the definition of Fox H-function, we get the PDF of $N$-aperture-based UWOC channel in \eqref{pdf_snr_mrc_uw_pe}.
	
	Now, we use \eqref{pdf_snr_mrc_uw_pe} in $F_{\gamma_N}(\gamma)=\int_{0}^{\gamma}f_{\gamma_N}(\gamma)d\gamma$ and apply the definition of Fox H-function to find the CDF of SNR in \eqref{cdf_snr_mrc_uw_pe}
\end{IEEEproof}
Denoting $\gamma_{\rm th}$ as the SNR threshold, the outage probability $P_{\rm out} P(\gamma \leq \gamma_{\rm th})$ is an important parameter for deducing the diversity order of the system. 	We can use the CDF of SNR in \eqref{cdf_snr_mrc_uw_pe} with $\gamma=\gamma_{\rm th }$ to get an exact expression of the outage probability over an EGG fading channel with pointing errors. In the following Lemma, we develop an asymptotic expression for the outage probability at a high SNR: 
\begin{my_lemma}	
The asymptotic expression of outage probability at high SNR $\gamma_0 \to \infty$ is given as:
	\begin{flalign}
	&P^{\infty}_{\rm out}=\prod_{i=1}^{N} \omega_i \sum_{k=1}^{2N}\frac{1}{\varepsilon_k}\frac{\prod_{j=1, j\neq k}^{N}\Gamma(E_j-E_k\frac{\varepsilon_j}{\varepsilon_k})\Gamma(\frac{E_k}{\varepsilon_k})}{\prod_{j=2}^{N+1}\Gamma(\mathcal{A}_j-E_k\frac{\alpha_j}{\varepsilon_k})\Gamma(1+\frac{E_k}{\varepsilon_k})}\nonumber\\& \left[\left(\frac{1}{\gamma_0\lambda^2_{i}A^2_{i}}\right)^{\frac{1}{N}}\gamma\right]^{\frac{E_k}{\varepsilon_k}} + \prod_{i=1}^{N} \frac{(1-\omega_i)}{ c_i \Gamma(a_i)} \sum_{k=1}^{2N}\frac{1}{\xi_k} \nonumber\\& \frac{\prod_{j=1, j\neq k}^{N}\Gamma(\mathcal{T}_j-\mathcal{T}_k\frac{\xi_j}{\xi_k})\Gamma(\frac{\mathcal{T}_k}{\xi_k})}{\prod_{j=2}^{N+1}\Gamma(\mathcal{Q}_j-\mathcal{T}_k\frac{\psi_j}{\xi_k})\Gamma(1+\frac{\mathcal{T}_k}{\xi_k})}\left[\left(\frac{1}{\gamma_0b^2_{i}A^2_{i}}\right)^{\frac{1}{N}}\gamma\right]^{\frac{\mathcal{T}_k}{\xi_k}}
	\label{cdf_snr_mrc_uw_pe_inf}
	\end{flalign}
	where $\mathcal{A}_j=\mathcal{A}_k=\{1, \rho_{1}^2+1, \cdots, \rho_{N}^2+1\}$, $\alpha_j=\alpha_k=\{1, \{\frac{2}{N}\}, \cdots, \{\frac{2}{N}\}_N\}$, $E_j=E_k=\{1,\cdots, 1_N, \rho_{1}^2, \cdots, \rho_{N}^2, 0\}$, $\varepsilon_j=\varepsilon_k=\{\{\frac{2}{N}\}, \cdots, \{\frac{2}{N}\}_N, \{\frac{2}{N}\}, \cdots, \{\frac{2}{N}\}_N, 1\}$, $\mathcal{Q}_j=\mathcal{Q}_k=\{1, \frac{\rho_{1}^2}{c_1}+1, \cdots, \frac{\rho_{N}^2}{c_N}+1\}$, $\psi_j=\psi_k=\{1, \frac{2}{c_1N}, \cdots, \frac{2}{c_NN}\}$, $\mathcal{T}_j=\mathcal{T}_k=\{a_1, \cdots, a_N, \frac{\rho_{1}^2}{c_1}, \cdots, \frac{\rho_{N}^2}{c_N}, 0\}$, and $\xi_j=\xi_k=\{\frac{2}{c_1N}, \cdots, \frac{2}{c_NN}, \frac{2}{c_1N}, \cdots, \frac{2}{c_NN}, 1\}$.

\end{my_lemma}
\begin{IEEEproof}
	We apply the identity \cite[eq. $1.8.4$]{Kilbas} on \eqref{cdf_snr_mrc_uw_pe} with $\gamma=\gamma_{\rm th }$ to find the asymptotic expression of the outage probability at high SNR $\gamma_0 \to \infty$.
\end{IEEEproof}

Combining the exponents of the SNR term in \eqref{cdf_snr_mrc_uw_pe_inf}, the diversity order can be obtained as $\text{DO}^{\rm MRC}=\min\{\sum_{i=1}^{N}\frac{1}{2}, \sum_{i=1}^{N}\frac{a_ic_i}{2}, \sum_{i=1}^{N}\frac{\rho_i^2}{2}\}$. For i.i.d underwater link condition i.e., $a_i=a$, $c_i=c$ and $\rho^2_i=\rho^2$,  $\forall i$, therefore $\text{DO}^{\rm MRC}=\min\{\frac{N}{2}, \frac{Nac}{2}, \frac{N\rho^2}{2}\}$. Thus, the  diversity order of an $N$-aperture system is improved by a factor $N$.
\subsection{Selection Combining}
Selection combining is a simple diversity scheme in which the received signal of the highest SNR is chosen for processing at the receiver. Using the SC scheme, CDF and PDF of the best SNR for i.i.d underwater link are given as $F_{\gamma_{\rm SC}}({\gamma}) = (F_\gamma({\gamma}))^N $ and $f_{\gamma_{\rm SC}}({\gamma}) = N(F_\gamma({\gamma}))^{N-1}f_\gamma({\gamma})$, respectively.

To develop the diversity order of the SC system, we present  asymptotic expressions for the PDF and CDF of SNR for an $N$ aperture-based UWOC system with SC diversity scheme over EGG turbulence channel with pointing errors in the following Lemma:

\begin{my_lemma}
An asymptotic expression for the CDF of SNR for $N$ aperture-based UWOC system with SC diversity scheme is given as: 
\begin{eqnarray}
&F_{\gamma_{\rm SC}}(\gamma) = \sum\limits_{k_1 = 0}^{N}\binom{N}{k_1} \Bigg[\sum\limits_{k_2 = 0}^{k_1}\binom{k_1}{k_2} \bigg(\frac{\Gamma(\rho^2-1)}{\Gamma(\rho^2)} \frac{\omega\rho^2}{\lambda A_0} \sqrt{\frac{\gamma}{\gamma_0}}\bigg)^{k_2}\nonumber\\& \bigg( \omega\Gamma(1-\rho^2)\left(\frac{1}{\lambda A_0} \left(\sqrt{\frac{\gamma}{\gamma_0}}\right)\right)^{\rho^2}\bigg)^{k_1-k_2}\Bigg] \Bigg[\sum\limits_{k_3 = 0}^{N-k_1}\nonumber\\& \binom{N-k_1}{k_3} \bigg(\frac{(1-\omega)\rho_i^2\Gamma(\frac{\rho^2}{c}-a)}{c\Gamma(\frac{\rho^2}{c}+1-a)\Gamma(1+a)} \left(\frac{1}{b A_0}\sqrt{\frac{\gamma}{\gamma_0}}\right)^{ac} \bigg)^{k_3}\nonumber\\&\bigg((1-\omega)\Gamma(a-\frac{\rho^2}{c})\left(\frac{1}{b A_0}\sqrt{\frac{\gamma}{\gamma_0}}\right)^{\rho^2}\bigg)^{N-k_1-k_3}\Bigg]
\label{cdf_uw_pe_approx_N_2}
\end{eqnarray}
\end{my_lemma}

\begin{IEEEproof}
Applying asymptotic series expansions \cite[eq. 07.34.06.0006.01]{Wolfram} of Meijer’s G function
\begin{eqnarray}
 &G_{2,3}^{2,1}\left(\begin{array}{c}1,\rho^2+1\\1,\rho^2,0\end{array}\left|\frac{1}{\lambda A_0} \left(\sqrt{\frac{\gamma}{\gamma_0}}\right)\right.\right) = \frac{\Gamma(\rho^2-1)}{\Gamma(\rho^2)} \frac{1}{\lambda A_0} \nonumber\\&\left(\sqrt{\frac{\gamma}{\gamma_0}}\right) + \frac{\Gamma(1-\rho^2)\Gamma(\rho^2)}{\Gamma(1+\rho^2)}\left(\frac{1}{\lambda A_0} \left(\sqrt{\frac{\gamma}{\gamma_0}}\right)\right)^{\rho^2}
 \end{eqnarray}
 and
 \begin{eqnarray} &G_{2,3}^{2,1}\left(\begin{array}{c}1,\frac{\rho^2}{c}+1\\a,\frac{\rho^2}{c},0\end{array}\left|\frac{1}{b^{c} A_0^{c}}\left(\sqrt{\frac{\gamma}{\gamma_0}}\right)^{c}\right.\right)= \frac{\Gamma(\frac{\rho^2}{c}-a)\Gamma(a)}{\Gamma(\frac{\rho^2}{c}+1-a)\Gamma(1+a)} \nonumber\\&\left(\frac{1}{b^{c} A_0^{c}}\left(\sqrt{\frac{\gamma}{\gamma_0}}\right)^{c}\right)^{a} +\frac{\Gamma(a-\frac{\rho^2}{c})\Gamma(\frac{\rho^2}{c})}{\Gamma(1+\frac{\rho^2}{c})}(\frac{1}{b^{c} A_0^{c}}\left(\sqrt{\frac{\gamma}{\gamma_0}}\right)^{c})^{\frac{\rho^2}{c}}
  \end{eqnarray}
in \eqref{cdf_uw_pe}, we get the asymptotic expansion of the CDF of SNR for a single-aperture UWOC system as:

\begin{eqnarray}
&F_\gamma(\gamma) = \omega\rho^2 \bigg(\frac{\Gamma(\rho^2-1)}{\Gamma(\rho^2)} \frac{1}{\lambda A_0} \left(\sqrt{\frac{\gamma}{\gamma_0}}\right)\nonumber\\& + \frac{\Gamma(1-\rho^2)\Gamma(\rho^2)}{\Gamma(1+\rho^2)}\left(\frac{1}{\lambda A_0} \left(\sqrt{\frac{\gamma}{\gamma_0}}\right)\right)^{\rho^2}\bigg)\nonumber\\& +  \frac{(1-\omega)\rho^2}{c \Gamma(a)} \bigg(\frac{\Gamma(\frac{\rho^2}{c}-a)\Gamma(a)}{\Gamma(\frac{\rho^2}{c}+1-a)\Gamma(1+a)} \left(\frac{1}{b^{c} A_0^{c}}\left(\sqrt{\frac{\gamma}{\gamma_0}}\right)^{c}\right)^{a} \nonumber\\&+\frac{\Gamma(a-\frac{\rho^2}{c})\Gamma(\frac{\rho^2}{c})}{\Gamma(1+\frac{\rho^2}{c})}(\frac{1}{b^{c} A_0^{c}}\left(\sqrt{\frac{\gamma}{\gamma_0}}\right)^{c})^{\frac{\rho^2}{c}}\bigg)
\label{cdf_uw_pe_approx}
\end{eqnarray}

To get the CDF of SNR for an $N$ aperture-based UWOC system ($F_{\gamma_{\rm SC}}({\gamma}) = (F_\gamma({\gamma}))^N $), we apply a binomial expansion $(a+b)^N = \sum_{k = 0}^{N}\binom{N}{k}a^k b^{N-k}$ in \eqref{cdf_uw_pe_approx}  to get \eqref{cdf_uw_pe_approx_N_2}.
\end{IEEEproof}
Similar to the MRC-based receiver, combining   the exponent of the SNR $\gamma_0$ from \eqref{cdf_uw_pe_approx_N_2}, the diversity order for the SC-receiver can be derived as $\text{DO}^{\rm SC}=\min\{\frac{N}{2}, \frac{Nac}{2}, \frac{N\rho^2}{2}\}$.

\begin{table}[t]	
	\renewcommand{\arraystretch}{01}
	\caption{Simulation Parameters}
	\label{table:simulation_parameters}
	\centering
	\begin{tabular}{|c|c|c|}
		\hline 	
		Transmitted optical power &$P_t$ & $-35$ to $20$ \mbox{dBm} \\ \hline
		
		AWGN variance &$\sigma_w^2$ & $10^{-14}~\rm {A^2/GHz}$ \\ \hline	
		Total link distance &$l$ & $50$ \mbox{m}\\ \hline
		Extinction coefficient& $\alpha$ & $0.056$ \\ \hline
		Pointing errors parameters& $A_0$ & \{$0.3900, 0.8532, 1$\} \\ &$\rho$ & \{$0.5718, 0.8863, 8$\}
		\\ \hline EGG distribution parameters \cite{Zedini2019}
		& $\omega$ &  $0.1770$\\& $\lambda$ &  $0.4687$ \\& $a$ & $0.6302$ \\& $b$ & $1.1780$\\& $c$ & $0.8444$	\\ \hline Threshold SNR & $\gamma_{\rm th}$& $60$ {\rm dB}\\ \hline
	\end{tabular}
	\label{Simulation_Parameters}	 
\end{table}
\section{Numerical and Simulation Results}
In this section, we demonstrate the outage probability performance of multi-aperture UWOC systems over EGG oceanic turbulence conditions with significant and negligible pointing errors using MRC and  SC diversity schemes. We also compare the outage performance of the multi-aperture UWOC with the single-aperture $(N = 1)$. We validate our derived analytical expressions using Monte-Carlo (MC) simulation (averaged over $10^7$ channel realizations). For the calculations of Meijer’s G and Fox’s H-function, we use standard inbuilt MATLAB and Mathematica libraries. The EGG oceanic turbulence channel and pointing errors parameters are obtained from laboratory experiments, as given in Table \ref{Simulation_Parameters}.

 \begin{figure}[tp]
	\centering
    \includegraphics[width=\columnwidth]{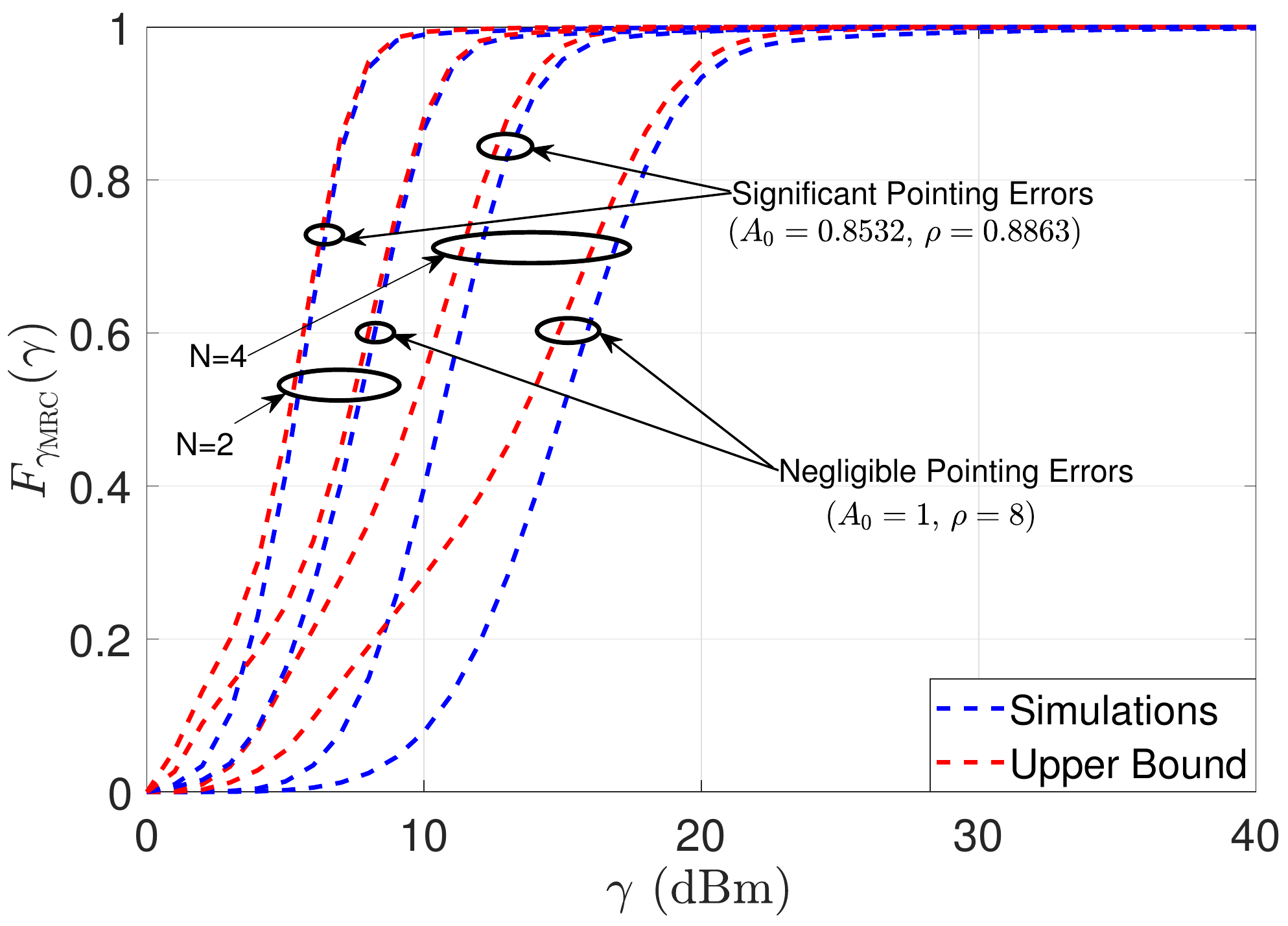}
	\caption{CDF of the SNR using MRC scheme.}
	\label{cdf_snr}
\end{figure}

\begin{figure}[tp]
		\centering
	\includegraphics[width=\columnwidth]{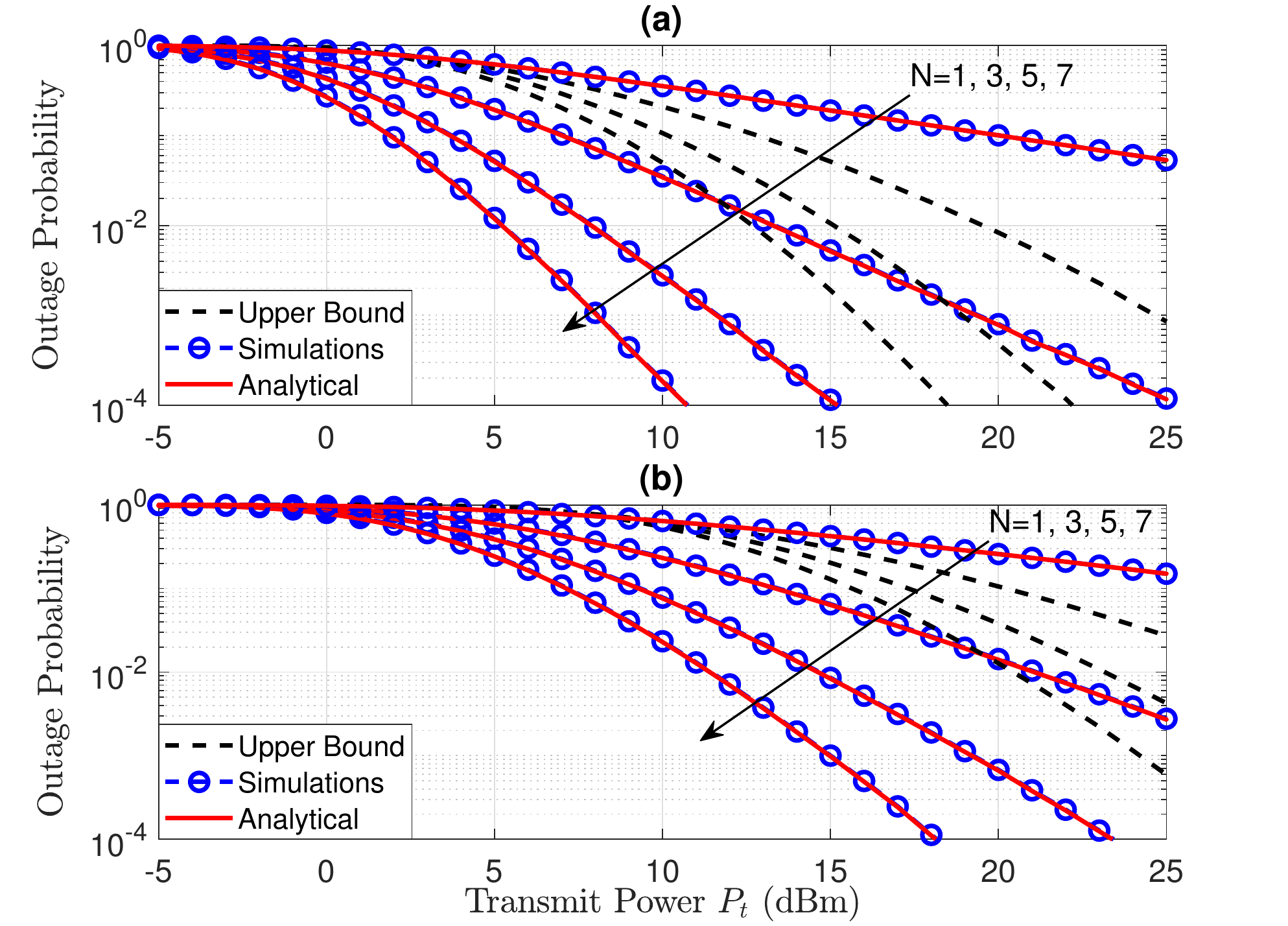}
	\caption{Outage probability performance of the UWOC system using MRC scheme, (a)  Significant pointing errors $A_0=0.8532$ and $\rho=0.8863$, (b) Negligible pointing errors $A_0=1$ and $\rho=8$.}
	\label{out_prob_mrc}
\end{figure}

\begin{figure}[tp]
		\centering
	\includegraphics[width=\columnwidth]{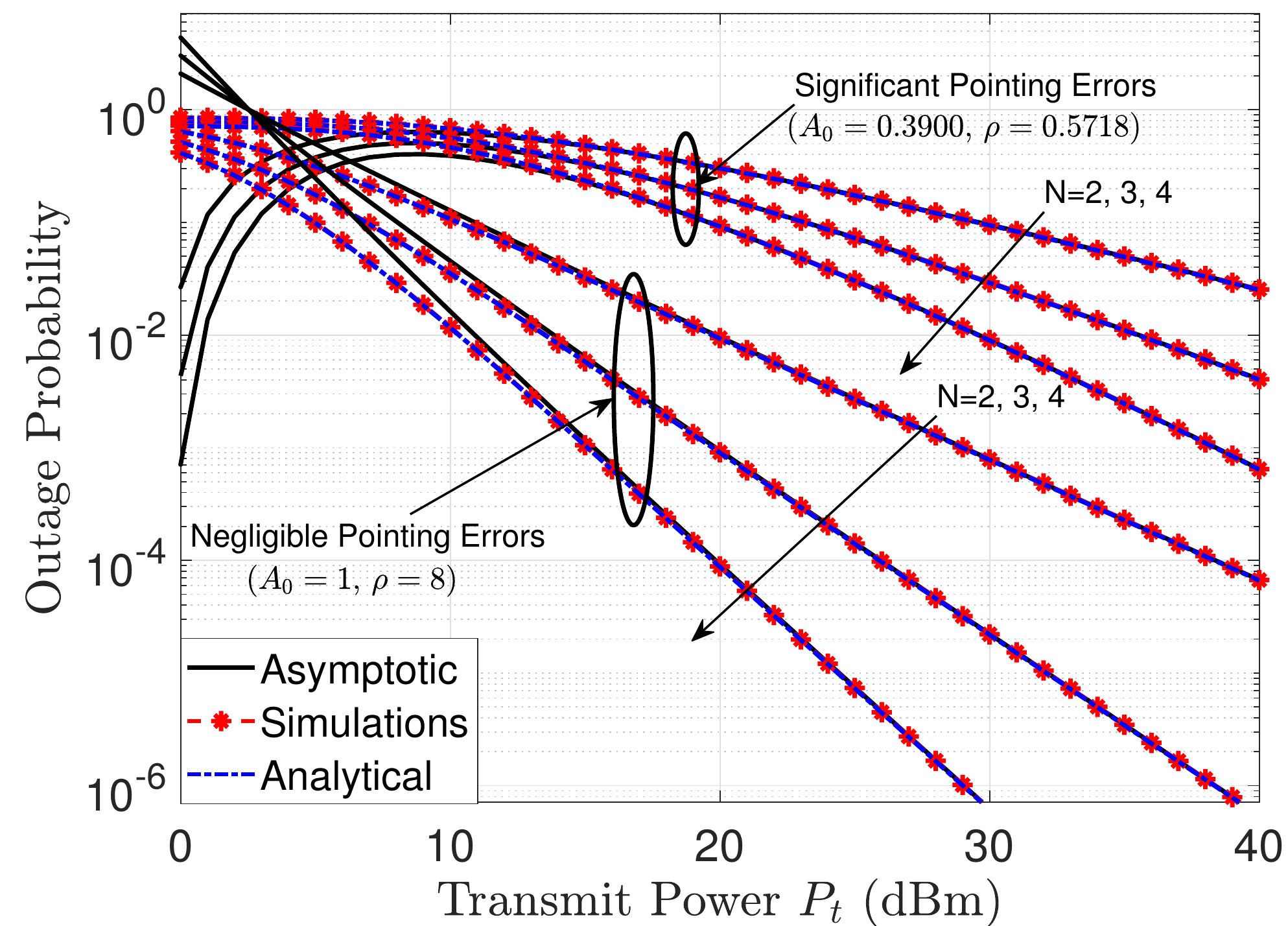}
	\caption{Outage probability performance of the UWOC system using SC scheme.}
	\label{out_prob_sc}
\end{figure}
First, we analyze the CDF of SNR in Fig.~\ref{cdf_snr} for the MRC diversity scheme. The figure shows the derived upper bound is close to the exact when the number of apertures is small. However, there exists a gap between the simulation and upper bound increases as we increase the number of received aperture $N$. It can also be seen that there is a difference between significant and negligible pointing errors case. 
Next, we demonstrate the outage probability of the UWOC system using the MRC scheme in Fig.~\ref{out_prob_mrc}. We consider $N=1, 3, 5, 7$ and pointing errors parameters $A_0=0.8532$ and $\rho=0.8863$. In this figure, we plot the outage probability performance for the single and multi-aperture systems along with the derived upper bound expression. It can be seen from the figure that the multi-aperture system provides better performance as compared to the single-aperture system.

Finally, we plot the outage probability versus transmit power for the SC scheme in Fig.~\ref{out_prob_sc}. We have used different values of $N=2, 3, 4$ and pointing errors parameters $A_0=0.3900$ and $\rho=0.5718$.  In this figure, we compare both the case with significant and negligible pointing errors along with the number of receiver antennas $N$. It can be seen from Fig.~\ref{out_prob_sc} that at high SNR, the asymptotic expression is almost matched with the analytical and simulation results.

It can also be seen  that the slope of the outage probability in Fig.~\ref{out_prob_mrc} and Fig.~\ref{out_prob_sc} confirms the diversity order of the system  $\min\{\frac{N}{2}, \frac{Nac}{2}, \frac{N\rho^2}{2}\}$ for the considered  multi-aperture UWOC systems.

\section{Conclusions}
In this paper, we presented the performance of a multi-aperture UWOC system under the combined effect of oceanic turbulence and pointing errors using MRC and SC diversity schemes at the receiver. After statistically characterizing the received instantaneous SNR at the multi-aperture-based receiver, we derived the expressions for the outage probability of both the combining schemes. Furthermore, we analytically obtained the considered system's diversity order by deriving the asymptotic expression of outage probability. Numerical analysis and Monte Carlo simulations show that the derived analytical expressions are close to each other at high SNR. It was also demonstrated that the multi-aperture UWOC system outperforms the single-aperture-based counterpart. Thus, the proposed multi-aperture model can be a promising alternative for high-speed UWOC links over diverse channel conditions.
	
\bibliographystyle{IEEEtran}
\bibliography{bib_file}

\end{document}